\documentclass[conference]{IEEEtran}
\IEEEoverridecommandlockouts

\usepackage[cmyk]{xcolor}
\usepackage[hidelinks]{hyperref}
\usepackage[utf8]{inputenc}
\usepackage{newtxtext,newtxmath}
\usepackage[american]{babel}
\usepackage{csquotes}
\usepackage[final]{microtype}
\usepackage{multirow}
\usepackage{xurl}
\usepackage{booktabs}
\usepackage{tabularx,ragged2e}
\newcolumntype{L}{>{\raggedright\arraybackslash}X}
\newcolumntype{R}{>{\raggedleft\arraybackslash}X}
\usepackage{threeparttable}
\usepackage{stmaryrd}
\usepackage{float}

\usepackage{pgfplots}
\pgfplotsset{compat=1.17}
\usepackage[outline]{contour}
\contourlength{0.2em}
\usepackage{tikz}
\usetikzlibrary{patterns}
\usepgfplotslibrary{fillbetween}
\usetikzlibrary{patterns}
\usetikzlibrary{positioning, calc}

\usepackage{siunitx}
\DeclareSIUnit\tx{tx}
\DeclareSIPrefix{\noop}{}{0} 
\usepackage[
  backend=biber,
  style=ieee
]{biblatex}
\usepackage{acro}
\DeclareAcronym{PoW}{
  short = PoW,
  long  = proof-of-work
}
\DeclareAcronym{PoS}{
  short = PoS,
  long  = proof-of-stake
}
\DeclareAcronym{DLT}{
  short = DLT,
  long  = distributed ledger technology
}
\DeclareAcronym{TPS}{
  short = tps,
  long  = transactions per second
}
\DeclareAcronym{LPoS}{
  short = LPoS,
  long  = liquid proof-of-stake
}
\DeclareAcronym{NPoS}{
  short = NPoS,
  long  = nominated proof-of-stake
}
\DeclareAcronym{PPoS}{
  short = PPoS,
  long  = pure proof-of-stake
}
\DeclareAcronym{CPU}{
  short = CPU,
  long  = central processing unit
}
\DeclareAcronym{HDD}{
  short = HDD,
  long  = hard disk drive
}
\DeclareAcronym{IOPS}{
  short = IOPS,
  long  = input/output operations per second
}
\DeclareAcronym{RAM}{
  short = RAM,
  long  = random-access memory
}
\DeclareAcronym{IO}{
  short = IO,
  long  = input/output
}
\DeclareAcronym{TDP}{
  short = TDP,
  long  = thermal design power
}
\DeclareAcronym{AWS}{
  short = AWS,
  long  = Amazon Web Services
}
\DeclareAcronym{EC2}{
  short = EC2,
  long  = Amazon Elastic Compute Cloud
}
\DeclareAcronym{L2}{
  short = L2,
  long  = layer 2
}
\DeclareAcronym{UTXO}{
  short = UTXO,
  long = unspent transaction output
}
\DeclareAcronym{zk}{
  short = zk,
  long  = zero-knowledge
}
\DeclareAcronym{DAG}{
  short = DAG,
  long  = directed acyclic graph
}

\usepackage[nomessages]{fp}

\newcommand\TpdVisa{150000000}
\FPeval{\TpsVisa}{\TpdVisa / (24 * 60 * 60)}
\FPeval{\AnnualisedConsumptionVisa}{706000} 
\FPeval{\PowerConsumptionVisa}{\AnnualisedConsumptionVisa * 1000000000 / (60 * 60 * 24 * 365.25)} %
\FPeval{\PerTxConsumptionVisa}{(\AnnualisedConsumptionVisa * 227.8) / (\TpsVisa * 60 * 60 * 24 * 365.25)} 

\newcommand\TpsBitcoin{2.62}
\FPeval{\AnnualisedConsumptionBitcoinLow}{29550000000} 
\FPeval{\AnnualisedConsumptionBitcoinHigh}{305010000000} 
\FPeval{\PowerConsumptionBitcoinLow}{\AnnualisedConsumptionBitcoinLow / (60 * 60 * 24 * 365.25)} 
\FPeval{\PowerConsumptionBitcoinHigh}{\AnnualisedConsumptionBitcoinHigh / (60 * 60 * 24 * 365.25)} 
\FPeval{\PerTxConsumptionBitcoinLow}{\AnnualisedConsumptionBitcoinLow / (\TpsBitcoin * 60 * 60 * 24 * 365.25)} 
\FPeval{\PerTxConsumptionBitcoinHigh}{\AnnualisedConsumptionBitcoinHigh / (\TpsBitcoin * 60 * 60 * 24 * 365.25)} 

\newcommand\NumValAlgoHistoric{1298}
\newcommand\NumValAlgo{1126}
\newcommand\NumValCardano{2958}
\newcommand\NumValEth{183753}
\newcommand\NumValPolka{297}
\newcommand\NumValPolkaHistoric{297}
\newcommand\NumValHedHistoric{20}
\newcommand\NumValHed{21}
\newcommand\NumValTezos{399}
\newcommand\NumValTezosHistoric{430}

\newcommand\TpsAlgoHistoric{11.5}
\newcommand\TpsAlgo{9.845}
\newcommand\TpsAlgoMax{1000}
\newcommand\TxCardano{157622}
\newcommand\PeriodCardano{432000}
\FPeval{\TpsCardano}{\TxCardano / \PeriodCardano}
\newcommand\TpsCardanoMax{257}
\newcommand\TpsPolka{0.12}
\newcommand\TpsPolkaMax{1000}
\newcommand\TpsHedHistoric{44.6}
\newcommand\TpsHed{48.2}
\newcommand\TpsHedMax{10000}
\newcommand\TpsEthLow{15.4}
\newcommand\TpsEthHigh{3000}
\newcommand\TpsEthMax{3000}
\newcommand\TpsTezos{1.7}
\newcommand\TpsTezosHistoric{0.4}
\newcommand\TpsTezosMax{40}

\newcommand\KappaAlgo{102.8}
\newcommand\LambdaAlgo{103.9}
\newcommand\KappaCardano{1267.8}
\newcommand\LambdaCardano{2959.2}
\newcommand\KappaCardanoPrecise{1267.763418}
\newcommand\LambdaCardanoPrecise{2959.188105}
\newcommand\KappaPolkadot{297.0}
\newcommand\LambdaPolkadot{0.0}
\newcommand\KappaHedera{7.6}
\newcommand\LambdaHedera{0.3}
\newcommand\KappaTezos{440.7}
\newcommand\LambdaTezos{-24.6}

\FPeval{\TECServer}{1473.5} 
\FPeval{\PServer}{(\TECServer * 1000) / 8766} 
\FPeval{\PSecServer}{(\PServer / 60 / 60) / 1000} 
\FPeval{\PHighEndServer}{328.0} 
\FPeval{\PSecHighEndServer}{(\PHighEndServer / 60 / 60) / 1000} 

\FPeval{\PRasPiLow}{3.4} 
\FPeval{\PRasPiHigh}{7.6} 
\FPeval{\PRasPiAvg}{(\PRasPiLow + \PRasPiHigh) / 2} 
\FPeval{\PSecRasPi}{(\PRasPiAvg / 60 / 60) / 1000} 

\FPeval{\PGAlgoOpt}{\NumValAlgo * \PRasPiAvg / 1000} 
\FPeval{\PGAlgoPess}{\NumValAlgo * \PServer / 1000} 
\FPeval{\PTxAlgo}{(\NumValAlgo * \PSecServer) / \TpsAlgo}
\FPeval{\PTxAlgoOpt}{(\NumValAlgo * \PSecRasPi) / \TpsAlgo}
\FPeval{\PTxAlgoMax}{(\NumValAlgo * \PSecServer) / \TpsAlgoMax}

\FPeval{\PGCardanoOpt}{\NumValCardano * \PRasPiAvg / 1000} 
\FPeval{\PGCardanoPess}{\NumValCardano * \PServer / 1000} 
\FPeval{\PGCardano}{\NumValCardano * \PSecServer * 60 * 60}
\FPeval{\PTxCardano}{(\NumValCardano * \PSecServer) / \TpsCardano}
\FPeval{\PTxCardanoOpt}{(\NumValCardano * \PSecRasPi) / \TpsCardano}
\FPeval{\PTxCardanoMax}{(\NumValCardano * \PSecServer) / \TpsCardanoMax}

\FPeval{\PGPolkaOpt}{\NumValPolka * \PRasPiAvg / 1000} 
\FPeval{\PGPolkaPess}{\NumValPolka * \PServer / 1000} 
\FPeval{\PGPolka}{\NumValPolka * \PSecServer * 60 * 60}
\FPeval{\PTxPolka}{(\NumValPolka * \PSecServer) / \TpsPolka}
\FPeval{\PTxPolkaOpt}{(\NumValPolka * \PSecRasPi) / \TpsPolka}
\FPeval{\PTxPolkaMax}{(\NumValPolka * \PSecServer) / \TpsPolkaMax}

\FPeval{\PGHedOpt}{\NumValHed * \PServer / 1000} 
\FPeval{\PGHedPess}{\NumValHed * \PHighEndServer / 1000} 
\FPeval{\PGHed}{\NumValHed * \PSecHighEndServer * 60 * 60}
\FPeval{\PTxHed}{(\NumValHed * \PSecHighEndServer) / \TpsHed}
\FPeval{\PTxHedOpt}{(\NumValHed * \PSecServer) / \TpsHed}
\FPeval{\PTxHedMax}{(\NumValHed * \PSecHighEndServer) / \TpsHedMax}

\FPeval{\PGEthOpt}{\NumValEth * \PRasPiAvg / 1000} 
\FPeval{\PGEthPess}{\NumValEth * \PServer / 1000} 
\FPeval{\PGEth}{\NumValEth * \PSecServer * 60 * 60}
\FPeval{\PTxEthLow}{(\NumValEth * \PSecServer) / \TpsEthLow}
\FPeval{\PTxEthLowOpt}{(\NumValEth * \PSecRasPi) / \TpsEthLow}
\FPeval{\PTxEthHigh}{(\NumValEth * \PSecServer) / \TpsEthHigh}
\FPeval{\PTxEthHighOpt}{(\NumValEth * \PSecRasPi) / \TpsEthHigh}
\FPeval{\PTxEthMax}{(\NumValEth * \PSecServer) / \TpsEthMax}

\FPeval{\PGTezosOpt}{\NumValTezos * \PRasPiAvg / 1000} 
\FPeval{\PGTezosPess}{\NumValTezos * \PServer / 1000} 
\FPeval{\PGTezos}{\NumValTezos * \PSecServer * 60 * 60}
\FPeval{\PTxTezos}{(\NumValTezos * \PSecServer) / \TpsTezos}
\FPeval{\PTxTezosOpt}{(\NumValTezos * \PSecRasPi) / \TpsTezos}
\FPeval{\PTxTezosMax}{(\NumValTezos * \PSecServer) / \TpsTezosMax}

\newif\ifdraft
\draftfalse
\ifdraft
\newcommand{\jsnote}[1]{ {\textcolor{orange} { ***Johannes: #1 }}}
\newcommand{\mpnote}[1]{ {\textcolor{blue} { ***Moritz: #1 }}}
\newcommand{\dpnote}[1]{ {\textcolor{brown} { ***Daniel: #1 }}}
\newcommand{\jinote}[1]{ {\textcolor{purple} { ***Juan: #1 }}}
\newcommand{\ptnote}[1]{ {\textcolor{green} { ***Paolo: #1 }}}
\newcommand{\nvnote}[1]{ {\textcolor{green} { ***Nikhil: #1 }}}
\newcommand{\jxnote}[1]{ {\textcolor{green} { ***Jiahua: #1 }}}

\else
\newcommand{\jsnote}[1]{}
\newcommand{\mpnote}[1]{}
\newcommand{\dpnote}[1]{}
\newcommand{\jinote}[1]{}
\newcommand{\ptnote}[1]{}
\newcommand{\nvnote}[1]{}
\newcommand{\jxnote}[1]{}
\fi

\DeclareRobustCommand*{\IEEEauthorrefmark}[1]{%
  \raisebox{0pt}[0pt][0pt]{\textsuperscript{\footnotesize #1}}%
}

\begin{document}

\title{The Energy Footprint of Blockchain Consensus Mechanisms Beyond Proof-of-Work}

\author{
    \IEEEauthorblockN{
        Moritz Platt\IEEEauthorrefmark{1}\IEEEauthorrefmark{2},
        Johannes Sedlmeir\IEEEauthorrefmark{3},
        Daniel Platt\IEEEauthorrefmark{4},
        Jiahua Xu\IEEEauthorrefmark{1},\\
        Paolo Tasca\IEEEauthorrefmark{1},
        Nikhil Vadgama\IEEEauthorrefmark{1},
        Juan Ignacio Ibañez\IEEEauthorrefmark{1}
    }

    \IEEEauthorblockA{\IEEEauthorrefmark{1}Centre for Blockchain Technologies, University College London, London, UK}
    \IEEEauthorblockA{\IEEEauthorrefmark{2}Department of Informatics, King's College London, London, UK}
    \IEEEauthorblockA{\IEEEauthorrefmark{3}FIM Research Center, University of Bayreuth, Bayreuth, Germany}
    \IEEEauthorblockA{\IEEEauthorrefmark{4}Department of Mathematics, Imperial College London, London, UK}
    \IEEEauthorblockA{moritz.platt@kcl.ac.uk,     \{jiahua.xu, p.tasca, nikhil.vadgama, j.ibanez\}@ucl.ac.uk
    \\johannes.sedlmeir@fim-rc.de,
    daniel.platt.17@ucl.ac.uk
}
}

\maketitle

\begin{abstract}
Popular permissionless \ac{DLT} systems using \ac{PoW} for Sybil attack resistance have extreme energy requirements, drawing stern criticism from academia, business, and the media.
\Ac{DLT} systems building on alternative consensus mechanisms, foremost \ac{PoS}, aim to address this downside.
In this paper, we take a first step towards comparing the energy requirements of such systems to understand whether they achieve this goal equally well.
While multiple studies have been undertaken that analyze the energy demands of individual blockchains, little comparative work has been done.
We approach this research gap by formalizing a basic consumption model for \ac{PoS} blockchains.
Applying this model to six archetypal blockchains generates three main findings:
First, we confirm the concerns around the energy footprint of \ac{PoW} by showing that Bitcoin's energy consumption exceeds the energy consumption of all \ac{PoS}-based systems analyzed by at least two orders of magnitude.
Second, we illustrate that there are significant differences in energy consumption among the PoS-based systems analyzed, with permissionless systems having an overall larger energy footprint.
Third, we point out that the type of hardware that validators use has a considerable impact on whether \ac{PoS} blockchains' energy consumption is comparable with or considerably larger than that of centralized, non-\Ac{DLT} systems.
\end{abstract}


\section{Introduction}
\label{sec:introduction}

In \acf{DLT} systems, consensus mechanisms fulfill multiple purposes surrounding the proposal, validation, propagation, and finalization of data~\cite{Xiao2020}.
A critical problem for \ac{DLT} systems are Sybil attacks in which an attacker creates an artificially large number of bogus identities~\cite{Douceur2002} to skew the results of majority decisions on the admission and order of transactions.
In permissioned networks, gatekeeping strategies can be applied that limit access to a network to previously vetted actors~\cite{Platt2021a}, thereby preventing such attacks.

However, for permissionless networks, in which participants can participate in consensus without any control~\cite{Tasca2019}, more complex mechanisms need to be applied to combat Sybil attacks.
\Acf{PoW} is an example of a Sybil attack resistance scheme that has been used in most early cryptocurrencies such as Bitcoin~\cite{Nakamoto2008}.
To counteract Sybil attacks, \ac{PoW} uses cryptographic puzzles of configurable difficulty with efficient verification such that it becomes computationally expensive for attackers to interfere with consensus~\cite{Back1997}.
However, by this design, the energy consumption of a \ac{PoW}-based cryptocurrency strongly correlates with its market capitalization, leading to an extreme energy demand for popular implementations~\cite{Sedlmeir2020}.
For instance, the electricity demand of Bitcoin is now in the same range as that of entire industrialized nations~\cite{Vries2018}.
Against this backdrop, many alternatives to \ac{PoW} have been proposed that do not rely on extensive computational effort~\cite{Ismail2019}.
Among those is \acf{PoS} in which participants with larger holdings of a cryptocurrency have larger influence in transaction validation.
While \ac{PoS} is generally understood as being more energy efficient than \ac{PoW}, the exact energy consumption characteristics of \ac{PoS}-based systems, and the influence that network throughput has on them, are not widely understood.

Two main approaches to quantify the energy consumption of a \ac{DLT} system have been assumed in the past.
One is to measure the consumption of a representative participant node and then extrapolate from this measurement.
An alternative approach is to develop a mathematical model that includes core metrics of a \ac{DLT} system to calculate its energy consumption.
Extensive research efforts have cumulated in best practices for determining the energy consumption of \ac{DLT} systems~\cite{Lei2021}.
So far, most work has focused on \ac{PoW} blockchains,\footnote{While not all distributed ledger technologies organize their data into chains of hash-linked blocks, the term \enquote{blockchain} is customarily interchangeable with \ac{DLT}. In the remainder, we follow this convention for simplicity.} and some research has investigated individual non-\ac{PoW} systems.
%
In this paper we propose a simple energy consumption model, applicable to a broad range of \ac{DLT} systems that use \ac{PoS} for Sybil attack resistance.
Specifically, this model considers the number of validator nodes, their energy consumption, and the network throughput based on which the energy consumption per transaction is estimated.
We present the results of applying this model to six \ac{PoS}-based systems. Our results illustrate that, while negligible compared to \ac{PoW}, the energy consumption of \ac{PoS} systems can still vary significantly.

The next section surveys related work in both experimental and mathematical models.
We then review selected \ac{PoS} systems -- Ethereum~2.0, Algorand, Cardano, Polkadot, Tezos, and Hedera -- and describe their relevant architectural features.
In the following section, we introduce our model in more detail and describe how the underlying data was obtained.
We apply the model to the systems selected, present the comparative results, and discuss limitations.
Finally, we conclude our study with potential avenues for future research.

\section{Related Work}
\label{sec:related-work}

We conducted an informal literature review using the search string \texttt{("Blockchain" OR "DLT" OR "Distributed Ledger") AND ("Energy Consumption" OR "Energy Demand" OR "Electricity Demand" OR "Carbon Footprint") year:[2008 TO *]}
on the Bielefeld Academic Search Engine (BASE).
We thereby obtained 413 results of various prior work on analyzing the energy demand of different \ac{DLT} systems, with a significant focus on \ac{PoW} blockchains in general and specifically Bitcoin.
Commonly, models take one of the following two forms.

\paragraph*{Experimental Models}

The first form revolves around conducting experiments using mining hardware and measuring its actual energy consumption, as done by Igumenov \textit{et al.} with different configurations of computational resources~\cite{Igumenov2019}. This approach has been used to derive consumption characteristics for different usage scenarios.
The \enquote{BCTMark} framework~\cite{Saingre2020}, for instance, allows for the deployment of an entire experiment stack, including the \ac{DLT} system under test.
Using load generators, a realistic network workload can be created.
The effects on the energy consumption of this setup under varying loads can subsequently be measured via energy sensors connected to the testbed.
An experimental study on the energy consumption of the non-\ac{PoW} XRP~ledger demonstrates that customizing validator hardware can yield reductions in energy demand~\cite{Roma2020}.
Metrics reported for common cryptocurrencies have been combined with testbed experiments to model the energy consumption behaviors of various consensus algorithms~\cite{Cole2018}.

\paragraph*{Mathematical Models}

An alternative method is to quantify assumptions about the environment in which a \ac{DLT} system operates.
Often, such models use a \enquote{top-down} approach that relies on publicly observable factors -- such as hash rate in the case of Bitcoin -- and associate them with common mining hardware or even seek to determine the hardware used via surveys~\cite{Lei2021}.
The papers of Gallersdörfer \textit{et al.}~\cite{Gallersdoerfer2020}, Küfeoglu and Özkuran~\cite{Kuefeoglu2019}, and Zade \textit{et al.}~\cite{Zade2019} are examples of this hash rate-based approach.
Sedlmeir~\textit{et al.}~\cite{Sedlmeir2020} undertake a basic comparison of different \ac{DLT} architectures with the conclusion that the energy consumption differs significantly depending on the design chosen.
A further study by the same authors~\cite{Sedlmeir2020a} refines previous models for Bitcoin's power consumption, such as the one by Vranken~\cite{Vranken2017}, and emphasizes that the driving forces behind power consumption are the Bitcoin price and the availability of cheap electricity.
Eshani~\textit{et al.}~\cite{Eshani2021} use a linear regression model to predict Ethereum's energy consumption based on the observed hash rate and difficulty level; however, the use of simplistic interpolation techniques alone is likely not an appropriate method for \ac{PoW} blockchains~\cite{Lei2021}.
Powell~\textit{et al.}~\cite{Powell2021} derive a mathematical model for the energy consumption of the \ac{PoS}-based Polkadot blockchain by extrapolating from the power demand of a single validator machine.

\section{Systems reviewed}

Our comparison set includes \ac{DLT} systems with high market capitalization that share a critical common denominator: using a \ac{PoS}-based consensus algorithm.
In \ac{PoS}, validators with a higher stake -- often in the form of the \ac{DLT} system's native currency -- influence the transaction validation more.
Thus, the scarce resource of energy to avoid Sybil attacks in \ac{PoW} is replaced by the scarce resource of capital in the cryptocurrency~\cite{Sedlmeir2020}.
Despite the commonalities, these systems differ in a range of other aspects, such as the minimum thresholds to validate and delegate, the necessity to lock-up tokens in order to stake (\enquote{bonding}), and the architecture of incentives consisting of penalties (\enquote{slashing}) and rewards beyond transaction fees (\enquote{block rewards}) (cf. \autoref{tab:protocols}).

When it comes to energy consumption, however, differences in the accounting model, transaction validation mechanism, and node permissioning setting~(cf. \autoref{tab:paradigms}), together with the architectural design of each system's specific \ac{PoS} protocol, are of particular relevance. In this section, we describe each of the \ac{PoS}-based systems with a focus on those aspects.
A full exploration of all possible factors is beyond the scope of this paper.

\paragraph*{Ethereum 2.0}
Ethereum is a highly popular permissionless blockchain that is currently transitioning from \ac{PoW} (Ethereum 1.0) to \ac{PoS} (Ethereum 2.0). In Ethereum~1.0, every full node needs to store all~350 GB of current state data.\footnote{\url{https://ethereum.org/en/developers/docs/storage/}} However, the storage of the full history of all transactions is used by archive nodes only. There are also light nodes storing only the header chains and requesting everything else from a full node on which they depend.\footnote{\url{https://ethereum.org/en/developers/docs/nodes-and-clients/}}
The sharding proposal (Ethereum 2.0 Phase 1), designed to limit compute, storage, and bandwidth needs, is not yet active.

\paragraph*{Algorand}
Algorand is a permissionless, account-based system where relay nodes store the entire ledger and non-relay nodes store approximately 1,000~blocks.
A proposal to limit storage needs through transaction expiration and sharding (\enquote{Vault}) is not yet active~\cite{Gilad2017}.

\paragraph*{Cardano}
Cardano is also permissionless and the only \ac{UTXO}-based system in our comparison set.
In Cardano, nodes store all transactions ever made. Its proposal for sidechains and sharding (\enquote{Basho}) is not yet active.
There is a probability of being selected as block-proposer for an epoch weighted by stake.
However, it is possible to delegate the stake to a stake pool, whose manager receives rewards when the pool is selected and then shares them with the delegators. Rewards are diminishing with the pool size if a pool is so large that it exceeds a saturation parameter. Non-selected stakers verify proposed blocks~\cite{Badertscher2018}.

\paragraph*{Polkadot}
In Polkadot's permissionless \ac{NPoS}, each node can delegate stake to up to 16 validators, among which the stake is always divided equally. Rewards to validators are proportionate to validation work, not to their stake.
Polkadot also distinguishes between archive nodes (storing all past blocks), full nodes (256 blocks), and light nodes (storing only runtime and current state, but no past blocks). The first five shards (\enquote{parachains}) have been already auctioned on the testnet \enquote{Kusama} but have not been deployed in the main chain.

\paragraph*{Tezos}
In Tezos' permissionless \ac{LPoS}, stake can also be delegated. Some delegates are block producers, other delegates verify; both receive rewards for it proportional to their stake~\cite{Goodman2014}.
Nodes have a \enquote{full mode} storing the necessary data needed to reconstruct the complete ledger state since the genesis block, but not contextual data from a checkpoint onwards; an \enquote{archive mode} where all blockchain data since the genesis block including contextual data such as past balances or staking rights beyond the checkpoints are stored; and \enquote{rolling mode} that only stores the minimal data that is necessary to validate blocks.

\paragraph*{Hedera}
In contrast to the other five systems studied,
Hedera is permissioned network that uses a \acf{DAG}-based data structure to store the transaction history (cf. \autoref{tab:paradigms}) and applies  \ac{PoS}~\cite{Baird2016}.
The network has its consensus nodes run solely by its council members at the moment, with the plan to open up to permissionless nodes in the future\footnote{\url{https://help.hedera.com/hc/en-us/articles/360000674017-Is-the-Hedera-public-network-permissioned-or-permissionless-}}. 
Transactions do not form blocks, but are spread through a \enquote{gossip about gossip} protocol where new information obtained by any node is spread exponentially fast though the network \cite{Hedera2021}.
The consensus calculation takes the form of a weighted average of all gossiping nodes' information such as transaction order, with the weight proportionate to a node's stake.

\begin{table}[!htb]
    \centering
    \begin{tabularx}{\columnwidth}{
    @{}
    X
    c
    c
    c
    c
    c
    c
    @{}}
        \toprule
        \multirow{2}{*}{\textbf{Platform}} & \multicolumn{2}{c}{\textbf{Accounting Model}} & \multicolumn{2}{c}{\textbf{Data structure}}
        & \multicolumn{2}{c}{\textbf{Permissioning}}
        \\
        & Account      & \ac{UTXO}      & Block     & \ac{DAG}   & P'ned & P'less \\ \midrule
        \textbf{Ethereum 2.0} & $\bullet$ &           & $\bullet$ &           &              & $\bullet$      \\
        \textbf{Algorand}     & $\bullet$ &           & $\bullet$ &           &              & $\bullet$      \\
        \textbf{Cardano}      &           & $\bullet$ & $\bullet$ &           &              & $\bullet$      \\
        \textbf{Polkadot}     & $\bullet$ &           & $\bullet$ &           &              & $\bullet$      \\
        \textbf{Tezos}        & $\bullet$ &           & $\bullet$ &           &              & $\bullet$      \\
        \textbf{Hedera}       & $\bullet$ &           &           & $\bullet$ & $\bullet$    &                \\ 
        \bottomrule
    \end{tabularx}
    \caption{Comparison of the analyzed DLT systems in accounting model, data structure, and node permissioning setting}
    \label{tab:paradigms}
\end{table}

\begin{table}[!htb]
    \centering
    \begin{tabularx}{\columnwidth}{Xccc}
        \toprule
        \textbf{Platform} & Bonding & Slashing & Rewards \\ \midrule
        \textbf{Ethereum 2.0} & Yes & Yes & Yes \\
        \textbf{Algorand} & No & No & Yes \\
        \textbf{Cardano} & No & No & Yes \\
        \textbf{Polkadot} & Yes & Yes* & Yes \\
        \textbf{Tezos} & No & Yes & Yes \\
        \textbf{Hedera} & No & No & No \\
        \bottomrule
    \end{tabularx}
    \caption{Properties of the PoS protocols used for Sybil attack resistance in the DLT systems analyzed
    }
    \label{tab:protocols}
\end{table}

\section{Method}
\label{sec:method}

Our model differs from previous work (cf. \autoref{sec:related-work}) in that we focus on the energy consumption per transaction, as opposed to the overall energy consumption of an entire system.
Nevertheless, existing models can be combined with additional data arising from the scientific literature, reports, and public ledger information to form a baseline that can be used to avoid time-consuming experimental validation.
Powell~\textit{et al.}~\cite{Powell2021} define an elementary mathematical model for the energy consumption of the Polkadot blockchain that can be generalized as
\begin{equation} \label{eq:powell}
p_t = p \cdot n_{\mathrm{val}},
\end{equation}
where $p_t$ is the overall average power the DLT system consumes, $p$ is average power consumed by a validator node, and $n_{\mathrm{val}}$ is the number of validator nodes.
Due to the comparatively low computational effort associated with \ac{PoS} and the intentionally relatively low throughput of permissionless blockchains to avoid centralization because of compute, bandwidth, or storage constraints~\cite{Buterin2021}, it can be assumed that validating nodes run on similar types of commodity server hardware, irrespective of the network load.

Under this assumption, the overall energy need of such a protocol is solely contingent on the number and hardware configuration of validator nodes.
In the context of this paper, we only consider the energy footprint of the consensus mechanisms itself.
We, therefore, only consider \textit{validators}\footnote{A node fulfilling this role goes by various names, e.g., \enquote{stake pool} for Cardano, or \enquote{baker} for Tezos.}, i.e., nodes that actively participate in a network's consensus mechanism by submitting and verifying the proofs necessary for Sybil attack resistance~\cite{Xiao2020}.
The overall number of nodes, including other \textit{full nodes} that replicate the transaction history without participating in consensus, is likely higher for all systems analyzed.
A key model parameter, therefore, is the number of validator machines running concurrently ($n_{\mathrm{val}}$).
This number can be established reliably, since it is stored on-chain as a key aspect of any \ac{PoS}-based protocol.
Table~\ref{tab:validators} shows the number of validators currently operating on each of the networks considered.

\begin{table}[h]
    \centering
    \sisetup{
        table-omit-exponent,
        table-number-alignment=right,
        table-auto-round=true}
    \begin{tabularx}{\columnwidth}{
        L
        @{\hspace{0.2cm}}
        S[table-format=6.0, fixed-exponent = 0]
        @{\hspace{0.2cm}}
        S[table-format=2.2, fixed-exponent = 0]
        @{\hspace{0.2cm}}
        S[table-format=5.0, fixed-exponent = 0]}
        \toprule
        \textbf{Platform} & \textbf{\# Validators} & \textbf{TPS Cont. (\si{\tx\per\second})} & \textbf{TPS Max. (\si{\tx\per\second})}  \\
        \midrule
        \textbf{Ethereum 2.0} & \NumValEth &  & \TpsEthMax \\
        \textbf{Algorand} &  \NumValAlgo & \TpsAlgo & \TpsAlgoMax \\
        \textbf{Cardano} &  \NumValCardano & \TpsCardano & \TpsCardanoMax \\
        \textbf{Polkadot} & \NumValPolka & \TpsPolka & \TpsPolkaMax \\
        \textbf{Tezos} & \NumValTezos & \TpsTezos & \TpsTezosMax \\
        \textbf{Hedera} & \NumValHed & \TpsHed & \TpsHedMax \\
        \bottomrule
    \end{tabularx}
    \vspace{0.1cm}
    \caption{The current number of validators, contemporary throughput, and the upper bound of throughput postulated (cf. Appendix \ref{appdx:validators}).}
    \label{tab:validators}
\end{table}

\paragraph*{Energy consumption per transaction}

To arrive at an energy consumption per transaction metric ($c_{\mathrm{tx}}$), the number of transactions per unit of time needs to be considered.
The actual numbers are dynamic and fluctuate over time.
The contemporary network throughput (\textit{Cont.}) is defined as the actual throughput a system experienced recently.
As a key metric, this can be derived from approximate timestamps that are associated with transactions on public ledgers.
The maximum postulated sustainable system throughput (\textit{Max.}) of the different protocols is derived from casual sources (cf. Appendix \ref{appdx:throughput}).
Note that these postulated figures are likely optimistic, that means, not necessarily reliable, as they originate not from controlled experiments but are anecdotal or come from promotional materials.
However, we consider these estimates acceptable as they have no direct influence on the energy consumption per transaction for a fixed contemporary network throughput.
They merely dictate the domain of the consumption function $f_{c_{\mathrm{tx}}}(l)$ that calculates the consumption per transaction depending on the overall system throughput~$l$ (measured in \si{\tx\per\second}).
Treating the average power consumed by a validator node ($p$, measured in~\si{\watt}) as a constant
means that an inverse relationship between consumption per transaction ($c_{\mathrm{tx}}$) and system throughput ($l$) can be established within the bounds of $(0, l_{\mathrm{max}}]$:
\begin{equation} \label{eq:model}
f_{c_{\mathrm{tx}}}(l) = \frac{n_{\mathrm{val}} \cdot p}{l}.
\end{equation}

\paragraph*{Modelling $c_{\mathrm{tx}}$ as a function of the number of transactions per second}

Equation~\eqref{eq:model} depends on two variables:
$n_{\mathrm{val}}$ and~$l$.
We will now present a model for $c_{\mathrm{tx}}$ that depends on one variable, namely $l$, only.
Data from the Cardano blockchain~\cite{Platt2021c} suggests that the number of validators $n_{\text{val}}$ and the number of transactions per second $l$ are positively correlated.
Namely, Pearson's correlation coefficient\footnote{The correlation coefficient takes values in $[-1,1]$ and a value of $\pm 1$ would imply that $n_{\text{val}}$ is an affine function in $l$.} for $n_{\text{val}}$ and $l$ for 375~data points from 29~Jul~2020 to 7~Aug~2021 is~$0.80$.
The correlation coefficient for $n_{\text{val}}$ delayed by 28~days and~$l$ (not delayed) for the same data is~$0.87$.
This is plausible for the following reason:
as the total number of users in a permissionless system increases, a share of these new users becomes validators and another non-disjoint share executes transactions, meaning that $n_{\text{val}}$ and $l$ are positively correlated.
For permissioned systems, it is still conceivable that the number of validators and throughput are linearly dependent and positively correlated because as new partner organizations are invited to run validator nodes, these partners may decide to use the system for their own applications, thereby increasing the number of transactions.
We also observe that in the case of Hedera, the number of validators and throughput are positively correlated:
the number of validator nodes has been continuously increasing; and throughput, while fluctuating from month to month, has increased year-to-year (cf. Appendix \ref{appdx:validators}).
Furthermore, it can be observed for the Algorand and Hedera blockchains that $n_{\text{val}}$~and~$l$ have increased from July to August~2021.
On the Polkadot blockchain, $n_{\text{val}}$ has remained constant from February to July~2021.
An exception is the Tezos blockchain for which $n_{\text{val}}$ has decreased while $l$ has increased from February to August 2021.
This trend has so far held true throughout the lifetime of the Tezos blockchain.
We note that in this case an affine function is not appropriate to model the dependence of $n_{\text{val}}$ on $l$, because $n_{\text{val}}$ would become negative for large values of $l$.
We will still compute the affine best approximation of $n_{\text{val}}$ in terms of $l$ for the Tezos blockchain, as it is an approximation of the first Taylor polynomial of $n_{\text{val}}$, and therefore a local model for $n_{\text{val}}$.

For simplicity we assume that the correlation is perfect, i.e., $n_{\text{val}} = \kappa+\lambda \cdot l$ for some $\kappa, \lambda \in \mathbb{R}, \lambda >0$, and using~\eqref{eq:model} we obtain
\begin{equation} \label{eq:model-substitute-nval}
f_{c_{\mathrm{tx}}}(l)
=
\frac{(\kappa+\lambda l) \cdot p}{l}.
\end{equation}

Because we could not obtain high-resolution historic data for Algorand, Polkadot, Tezos, and Hedera, we will later on compute $\kappa,\lambda$ based on two data points.
For Cardano, we use linear regression implemented as ordinary least squares regression to compute $\kappa,\lambda$ that have the maximum likelihood of modelling $f_{c_{\mathrm{tx}}}(l)$ under the assumption that $f_{c_{\mathrm{tx}}}(l)$ is an affine function with Gaussian noise.
The resulting values for $\kappa,\lambda$ can be found in Table~\ref{tab:kappa-lambda-values}.

\begin{table}[hbt]
    \centering
    \begin{tabularx}{\columnwidth}{
        X
        r
        r
    }
        \toprule
        \textbf{Platform} & {$\kappa$} & {$\lambda$}  \\
        \midrule
        \textbf{Algorand} &  \KappaAlgo & \LambdaAlgo \\
        \textbf{Cardano} &  \KappaCardano & \LambdaCardano \\
        \textbf{Polkadot} &  \KappaPolkadot & \LambdaPolkadot \\
        \textbf{Tezos} &  \KappaTezos & \LambdaTezos \\
        \textbf{Hedera} &  \KappaHedera & \LambdaHedera \\
        \bottomrule
    \end {tabularx}
    \vspace{0.1cm}
    \caption{Estimates for $\kappa,\lambda$ for different DLT platforms used in \eqref{eq:model-substitute-nval} to model the number of validators depending on the number of transactions per second.}
    \label{tab:kappa-lambda-values}
\end{table}

\paragraph*{Hardware Type and Compute Resource Utilization Considerations}

In stark contrast to energy-intensive \ac{PoW} systems, in \ac{PoS}, the computational effort relating to the participation in the consensus protocol can practically be considered independent of extraneous factors like cryptocurrency capitalization.
Numerous factors influence the overall energy consumption of a server with \ac{CPU} activity, hard disk drive operations and cooling contributing most significantly to it~\cite{Jaiantilal2010}.
Consensus-related energy demand in \ac{PoS} is generally constant, meaning it occurs irrespective of system load~\cite{Jaiantilal2010}.
Energy demand relating to \ac{CPU} time and input/output operations is, however, highly load-dependent~\cite{Zhou2019}.
Therefore, a realistic energy consumption estimate for a validator node needs to factor in both the minimum hardware requirement (i.e., how many \ac{CPU} cores or what amount of memory is required) as well as the utilization of that hardware.

Since it is close to impossible to determine which type of hardware is used by validators in actuality, we use an approximation derived from industry recommendations.
For permissionless systems and permissioned systems dramatically different hardware recommendations are put forward. 
The permissionless systems analyzed in this study, all traditional blockchains with comparatively large numbers of validators running full nodes that verify every transaction~\cite{Buterin2021}, demand comparatively low-powered hardware.
Hedera, the only permissioned system analyzed here, constitutes a high-\ac{TPS} system.
Such systems are characterized by a small number of nodes maintains consensus~\cite{Buterin2021}.
As such, the network performance is determined by the lowest-performing validator node\footnote{\url{https://docs.hedera.com/guides/mainnet/mainnet-nodes/node-requirements}}. 
Therefore, in order to achieve the postulated maximum throughput values, highly performant server hardware is demanded by the network operator.
We assumed that similar high-\ac{TPS} systems would have energy requirements in the same range.
This explains the difference in energy consumption per validator node between Hedera and the other traditional Blockchain systems. 

\begin{table}[h]
    \centering
    \sisetup{
        table-omit-exponent,
        table-number-alignment=right,
        table-auto-round=true}
    \begin{tabularx}{\columnwidth}{
        l
        L
        L
        S[table-format=3.1, fixed-exponent = 0]}
        \toprule
        \textbf{Config.} & \textbf{Hardware Type} & \textbf{Exemplar} & \textbf{Demand (\si{\watt})}  \\
        \midrule
        \textbf{Minimum} &  Small single-board computer & Raspberry Pi 4 & \PRasPiAvg \\
        \textbf{Medium} & General-purpose rackmount server & Dell PowerEdge R730 & \PServer \\
        \textbf{Maximum} & High-performance server & Hewlett Packard Enterprise ProLiant ML350 Gen10 & \PHighEndServer \\
        \bottomrule
    \end {tabularx}
    \vspace{0.1cm}
    \caption{Conceivable upper and lower bounds for the power demand of a validator machine}
    \label{tab:hardware-configurations}
\end{table}

To capture the uncertainty regarding appropriate hardware and expected hardware utilization in the model, three different validator configurations are considered (cf. Table~\ref{tab:hardware-configurations}):
a single-board computer, a general-purpose rackmount server for midsize and large enterprises, and a high-performance server.
For all configurations, hardware utilization based on typical workloads is assumed (cf. Appendix \ref{appdx:consumption}).
For traditional blockchains, we assume a power demand in the minimum to medium range (\SIrange[round-mode=places, round-precision=1]{\PRasPiAvg}{\PServer}{\watt}).
For high-\ac{TPS} systems, the medium to maximum range (\SIrange[round-mode=places, round-precision=1]{\PServer}{\PHighEndServer}{\watt}) is assumed.

\section{Results}

Table~\ref{tab:results} illustrates the application of the models described in~\eqref{eq:model} and~\eqref{eq:model-substitute-nval}
to estimate the energy consumption of the protocols considered under contemporary throughput, i.e., based on recent throughput measures (cf. Section~\ref{sec:method}).
To facilitate a broad overview, we also provide the global system-wide consumption of each \ac{DLT} system according to the model.
Furthermore, the table presents two estimates for energy consumption per \ac{DLT} system:
an \textit{optimistic} estimate assuming validator nodes are operated on the lower bound of the system range and a \textit{pessimistic} estimate that assumes validators utilize hardware on the higher bound (cf. Table~\ref{tab:hardware-configurations}).
As the merge of Ethereum mainnet with the beacon chain is outstanding, no contemporary throughput figures for Ethereum~2.0 can be established.
Instead, a broad projection between a lower bound, the current throughput of the Ethereum blockchain (\SI{\TpsEthLow}{\tx\per\second}), and an upper bound, the postulated maximum value following the merge (\SI{\TpsEthHigh}{\tx\per\second}), is presented (cf. Appendix~\ref{appdx:throughput}).

\begin{figure*}
    \centering
    \begin{tikzpicture}
        \begin{axis}[
            ymode = log,
            xmode = log,
            clip=false,
            width=0.85\textwidth,
            height=0.45\textwidth,
            xlabel = Throughput {[\si{\tx\per\second}]},
            xmin = 0.02, xmax = 11000,
            ylabel = Energy Consumption {[\si{\kWh\per\tx}]},
            ymax = 10000,
        ]


            \addplot[color=cyan!30, name path=AlgoServer, -|, line width=0.25mm, domain=0.02:1000]   {(\KappaAlgo + \LambdaAlgo * x) * \PSecServer / x};
            \addplot[color=cyan!30, name path=AlgoRasPi, -|, line width=0.25mm, domain=0.02:1000] {(\KappaAlgo + \LambdaAlgo * x) * \PSecRasPi / x};
            \addplot[pattern color=cyan!20, pattern=north east lines] fill between[of=AlgoServer and AlgoRasPi];

            \addplot[color=red!30, name path=TezosServer, |-|, line width=0.25mm, domain=0.437340278:1.7]   {(\KappaTezos + \LambdaTezos * x) * \PSecServer / x};
            \addplot[color=red!30, name path=TezosRasPi, |-|, line width=0.25mm, domain=0.437340278:1.7] {(\KappaTezos + \LambdaTezos * x) * \PSecRasPi / x};
            \addplot[pattern color=red!20, pattern=north east lines] fill between[of=TezosServer and TezosRasPi];

            \addplot[color=purple!30, name path=CardanoServer, -|, line width=0.25mm, domain=0.02:\TpsCardanoMax] {\KappaCardanoPrecise + \LambdaCardanoPrecise * x) * \PSecServer / x};
            \addplot[color=purple!30, name path=CardanoRasPi, -|, line width=0.25mm, domain=0.02:\TpsCardanoMax] {\KappaCardanoPrecise + \LambdaCardanoPrecise * x) * \PSecRasPi / x};
            \addplot[pattern color=purple!20, pattern=north west lines]fill between[of=CardanoServer and CardanoRasPi];

            \addplot[color=orange!30, name path=PolkadotServer, -|, line width=0.25mm, domain=0.02:1000]   {(\KappaPolkadot + \LambdaPolkadot * x) * \PSecServer / x};
            \addplot[color=orange!30, name path=PolkadotRasPi, -|, line width=0.25mm, domain=0.02:1000] {(\KappaPolkadot + \LambdaPolkadot * x) * \PSecRasPi / x};
            \addplot[pattern color=orange!20, pattern=north west lines] fill between[of=PolkadotServer and PolkadotRasPi];

            \addplot[color=green!30, name path=HederaServer, -|, line width=0.25mm, domain=0.02:10000]   {(\KappaHedera + \LambdaHedera * x) * \PSecHighEndServer / x};
            \addplot[color=green!30, name path=HederaRasPi, -|, line width=0.25mm, domain=0.02:10000] {(\KappaHedera + \LambdaHedera * x) * \PSecServer / x};
            \addplot[pattern color=green!20, pattern=north east lines] fill between[of=HederaServer and HederaRasPi];

            \node[circle, fill, scale=0.3, label={right:\scriptsize \bfseries\sffamily{\num{\PerTxConsumptionBitcoinLow}}}] at (axis cs:{\TpsBitcoin},{\PerTxConsumptionBitcoinLow}) {};
            \node[circle, fill, scale=0.3, label={right:\scriptsize \bfseries\sffamily{\num{\PerTxConsumptionBitcoinHigh}}}] at (axis cs:{\TpsBitcoin},{\PerTxConsumptionBitcoinHigh}) {};
            \draw (axis cs:{\TpsBitcoin},{\PerTxConsumptionBitcoinLow}) -- (axis cs:{\TpsBitcoin},{\PerTxConsumptionBitcoinHigh}) node [midway, label=left:{\small Bitcoin}] {};

            \node[circle, fill, scale=0.3, label={right:\scriptsize \bfseries\sffamily{\num{\PerTxConsumptionVisa}}}, label={left:\small VisaNet}] at (axis cs:{\TpsVisa},{\PerTxConsumptionVisa}) {};


            \node[color=cyan!30, circle, fill, scale=0.3, label={[font=\scriptsize,text=cyan] right:\bfseries\sffamily{\num{\PTxAlgo}}}] at (axis cs:{\TpsAlgo},{\PTxAlgo}) {};
            \node[color=cyan!30, circle, fill, scale=0.3, label={[font=\scriptsize,text=cyan] right:\bfseries\sffamily{\num{\PTxAlgoOpt}}}] at (axis cs:{\TpsAlgo},{\PTxAlgoOpt}) {};
            \draw[cyan!30] (axis cs:{\TpsAlgo},{\PTxAlgoOpt}) -- (axis cs:{\TpsAlgo},{\PTxAlgo}) node [midway, label={[font=\small,text=cyan] right:Algorand}] {};

            \node[color=red!30, circle, fill, scale=0.3, label={[font=\scriptsize,text=red] right:\bfseries\sffamily{\num{\PTxTezos}}}] at (axis cs:{\TpsTezos},{\PTxTezos}) {};
            \node[color=red!30, circle, fill, scale=0.3, label={[font=\scriptsize,text=red] right:\bfseries\sffamily{\num{\PTxTezosOpt}}}] at (axis cs:{\TpsTezos},{\PTxTezosOpt}) {};
            \draw[red!30] (axis cs:{\TpsTezos},{\PTxTezosOpt}) -- (axis cs:{\TpsTezos},{\PTxTezos}) node [midway, label={[font=\small,text=red] right:Tezos}] {};

            \draw[purple!30] (axis cs:{\TpsCardano},{\PTxCardanoOpt}) -- (axis cs:{\TpsCardano},{\PTxCardano}) node [midway, label={[font=\small,text=purple] right:Cardano}] {};
            \node[color=purple!30, circle, fill, scale=0.3, label={[font=\scriptsize,text=purple] right:\bfseries\sffamily{\num{\PTxCardano}}}] at (axis cs:{\TpsCardano},{\PTxCardano}) {};
            \node[color=purple!30, circle, fill, scale=0.3, label={[font=\scriptsize,text=purple] right:\bfseries\sffamily{\num{\PTxCardanoOpt}}}] at (axis cs:{\TpsCardano},{\PTxCardanoOpt}) {};

            \node[color=orange!30, circle, fill, scale=0.3, label={[font=\scriptsize,text=orange] left:\bfseries\sffamily{\num{\PTxPolka}}}] at (axis cs:{\TpsPolka},{\PTxPolka}) {};
            \node[color=orange!30, circle, fill, scale=0.3, label={[font=\scriptsize,text=orange] left:\bfseries\sffamily{\num{\PTxPolkaOpt}}}] at (axis cs:{\TpsPolka},{\PTxPolkaOpt}) {};
            \draw[orange!30] (axis cs:{\TpsPolka},{\PTxPolkaOpt}) -- (axis cs:{\TpsPolka},{\PTxPolka}) node [midway, label={[font=\small,text=orange] left:Polkadot}] {};

            \node[color=green!30, circle, fill, scale=0.3, label={[font=\scriptsize,text=green] right:\bfseries\sffamily{\num{\PTxHed}}}] at (axis cs:{\TpsHed},{\PTxHed}) {};
            \node[color=green!30, circle, fill, scale=0.3, label={[font=\scriptsize,text=green] right:\bfseries\sffamily{\num{\PTxHedOpt}}}] at (axis cs:{\TpsHed},{\PTxHedOpt}) {};
            \draw[green!30] (axis cs:{\TpsHed},{\PTxHedOpt}) -- (axis cs:{\TpsHed},{\PTxHed}) node [midway, label={[font=\small,text=green] left:Hedera Hashgraph}] {};
        \end{axis}
    \end{tikzpicture}
    \caption{The energy consumption per transaction is close to inversely correlated with throughput. For each system, the lower mark indicates the energy consumption under an optimistic validator hardware assumption while the upper mark indicates a pessimistic model. The consumption figures for Bitcoin and VisaNet are plotted for comparison (cf. Appendix~\ref{appdx:benchmark}). For Ethereum 2.0, no throughput metrics are available.}
    \label{fig:results}
\end{figure*}
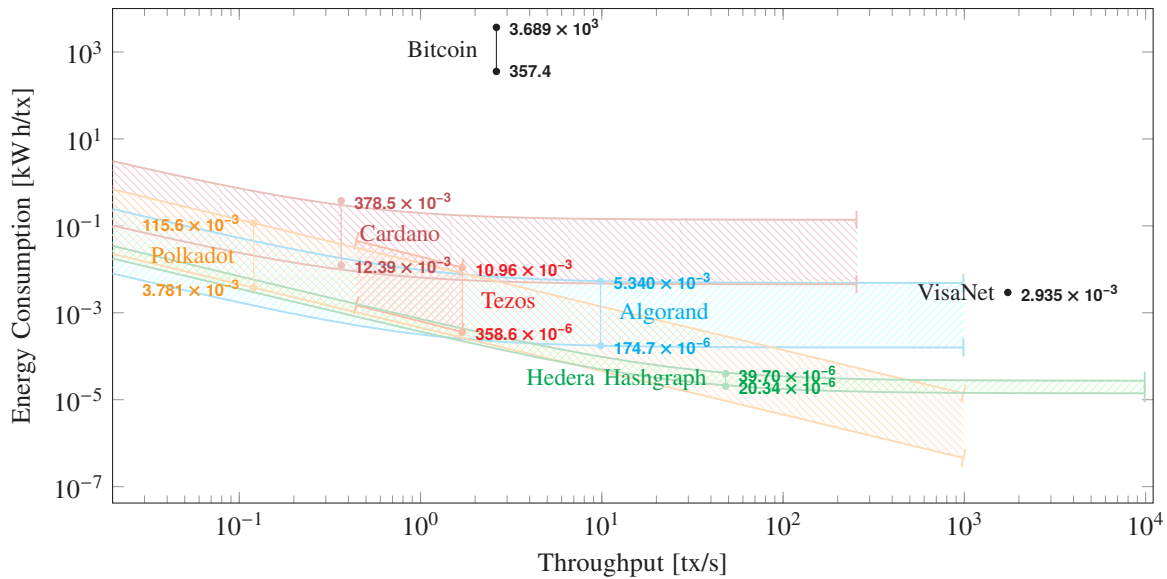

All estimates are based on the validator counts established earlier (cf.~~Section~\ref{sec:method}).
The plot of the model function shown in Figure~\ref{fig:results} visualizes the inverse relationship described earlier within the boundaries of the postulated throughput values (cf. Table~\ref{tab:validators}).
It also provides a projection of energy consumption as a function of system load, based on the model presented earlier which predicts the number of validators as a function of system load.
This projection is equally illustrated within the boundaries of the postulated throughput values, except in the case of the Tezos Blockchain for which no global model could be derived.

Based on this data, we can compare the energy consumption per transaction on two related systems:
first the \ac{PoW} cryptocurrency Bitcoin and second the VisaNet payment network (Figure~\ref{fig:results}).
It becomes evident that the consumption of Bitcoin -- overall and per-transaction -- is at least three orders of magnitude higher than that of the highest consuming \ac{PoS} system even under the most favorable assumptions.
While the difference between \ac{PoS} systems and VisaNet is less pronounced, it is evident that most of the former undercut the energy consumption of VisaNet in most configurations.

\begin{table}[ht]
    \centering
    \begin{threeparttable}[b]
        \sisetup{
            table-alignment=right,
            table-number-alignment=right,
            table-auto-round=true}
        \begin{tabularx}{\columnwidth}{
            L
            S[scientific-notation=false, table-format = 7.1]
            @{\hspace{1ex}}
            c
            @{\hspace{1ex}}
            S[scientific-notation=false, table-format = 8.1]
            S[scientific-notation=false, table-format = 4.5]
            @{\hspace{1ex}}
            c
            @{\hspace{1ex}}
            S[scientific-notation=false, table-format = 4.5]}
            \toprule
            \textbf{Platform} & \multicolumn{3}{c}{\textbf{Global (\si{\kilo\watt})}} & \multicolumn{3}{c}{\textbf{Per transaction (\si{\kilo\watt\hour\per\tx})}} \\
            \midrule
            \textbf{Eth. 2.0}\tnote{$\shortuparrow$} & \PGEthOpt & -- & \PGEthPess & \PTxEthHighOpt & -- & \PTxEthHigh \\
            \textbf{Eth. 2.0}\tnote{$\shortdownarrow$} & \PGEthOpt & -- & \PGEthPess & \PTxEthLowOpt & -- & \PTxEthLow \\
            \textbf{Algorand} & \PGAlgoOpt & -- & \PGAlgoPess & \PTxAlgoOpt & -- & \PTxAlgo \\
            \textbf{Cardano} & \PGCardanoOpt & -- & \PGCardanoPess & \PTxCardanoOpt & -- & \PTxCardano \\
            \textbf{Polkadot} & \PGPolkaOpt & -- & \PGPolkaPess & \PTxPolkaOpt & -- & \PTxPolka \\
            \textbf{Tezos} & \PGTezosOpt & -- & \PGTezosPess & \PTxTezosOpt & -- & \PTxTezos \\
            \textbf{Hedera} & \PGHedOpt & -- & \PGHedPess & \PTxHedOpt & -- & \PTxHed \\
            \midrule
            \textbf{Bitcoin} & 3373287.671 & -- & 34817351.598 & 360.393 & -- & 3691.407 \\
            \textbf{VisaNet} & \multicolumn{2}{c}{ } & 22387.1131 & \multicolumn{2}{c}{} & 0.0035819381 \\
            \bottomrule
        \end{tabularx}
        \begin{tablenotes}
            \item [$\shortuparrow$] High throughput projection
            \item [$\shortdownarrow$] Low throughput projection
        \end{tablenotes}
    \end{threeparttable}
    \caption{Global power consumption (i.e. the network-wide consumption of the DLT systems under consideration and VisaNet) and the energy consumed per transaction for contemporary throughput (see~Table~\ref{tab:validators})}
    \label{tab:results}
\end{table}

Pronounced differences between \ac{PoS}-based systems are equally evident from the results.
We observe low energy demand per transaction in active permissioned \ac{DLT} systems that are characterized by comparatively small numbers of validators and high throughput.
Less active permissionless systems show a higher energy demand per transaction due to comparatively lower throughput and a high number of validators.
This illustrates that not only for \ac{PoW}~\cite{Carter2021} but also for \ac{PoS} blockchains, \enquote{energy consumption per transaction} should not be the only metric considered for assessing the sustainability.
Particularly when utility is not approximately proportional to throughput, total energy consumption may be a more appropriate key figure.

\section{Discussion}

\subsection{Interpretations}

These results can primarily be understood as a clear confirmation of the common opinion that the energy consumption of \ac{PoW} systems, especially Bitcoin, is excessive.
Therefore, they can be interpreted as a strong argument for the modernization of \ac{PoW}-based systems towards \ac{PoS}.
Ethereum is taking a commendable lead in this respect with the development of Ethereum 2.0.
Furthermore, the results indicate that the energy consumption of different non-\ac{PoW} blockchains is surprisingly divergent (e.g., by a factor of about \num{1000} between the \ac{PoS} system with the highest consumption and the one with the lowest). 
In absolute terms, however, the consumption rates of \ac{PoS}-based systems are moderate and thus also much closer to the figures for traditional, centralized payment systems such as VisaNet.

The main reason why our model yields considerable divergence between \ac{PoS} systems is the different number of validators.
Specifically, in permissioned systems, energy consumption can be controlled through the ability to limit the number of validators on a network, so the permissioned network analyzed in this study is characterized by low energy consumption.
However, this observation does not warrant conclusions such as that permissioned systems are necessarily less energy consumptive.
Moreover, while in permissioned systems an operator can influence the number of nodes, it does not necessarily mean that that number must be lower. 

Even if a reducing effect of permissioning on energy consumption could be stated with certainty, this should not be misinterpreted as an argument for increased centralization or an argument for permissioned networks over permissionless ones.
This becomes obvious when considering a permissioned \ac{DLT} system in extremis:
such a system would consist of only a single validator node and would thus be effectively centralized.
This hypothetical scenario shows that, if a permissioned paradigm is applied, close attention should be paid to system entry barriers enforced through gatekeeping capabilities.
If not, there is a risk of centralization, which may offer advantages in terms of energy consumption, but will negate the functional advantages of a decentralized paradigm.
Of practical relevance is also the result that the selection of suitable validator hardware is central to energy consumption.
Information regarding adequate hardware for validators is often inconsistent.
Therefore, standardized recommendations should be put forward to help operators of validator nodes in selecting the most energy-efficient hardware configuration.

This study is only a first step towards quantifying the energy consumption of \ac{PoS} systems.
However, despite its limitations, it gives impetus to designers of decentralized systems by revealing the dependency between validator number, load, and hardware configuration.
Our model can thus be used to determine the carbon footprint of a particular use case.
It can furthermore prompt operators of validator nodes to carefully select suitable hardware.

\subsection{Limitations}

So far, we have used broad consumption ranges to model the energy consumption of individual validator nodes.
While we are confident that the actual energy consumption is in fact within these ranges, underlying characteristics of different \ac{PoS} protocols that might impact energy consumption, such as the accounting model, have been ignored.
Second, while assuming that the electricity consumption of a validator node is independent of system throughput is well justified for the permissionless systems analyzed~\cite{Buterin2021}, permissioned systems that are designed to support high throughput may not warrant such assumption.
While we have accounted for this by assuming more powerful hardware for permissionless high\ac{TPS} systems, more work is needed to understand permissioned blockchains' energy consumption characteristics better.
Moreover, the impact of different workloads on energy consumption should be considered; for example, simple payments transactions may have lower computational requirements when compared to other smart contract calls, but so far we have not distinguished between transaction types.

Further, while our model suggests that \ac{PoS} systems can remain energy-efficient while scaling up to VisaNet throughput levels, there is no hard evidence in support of this argument, as no \ac{DLT}-based system has experienced a sustained volume of this magnitude to date on the base level.

We ignored the possibility of achieving effectively higher throughput than the specified maximum through \ac{L2} solutions, such as the Lightning network or via optimistic and \ac{zk}-rollups that are receiving increasing attention.

Finally, although there are reasons to support its plausibility, the assumption that an affine function can be used to express the number of validators in terms of throughput is questionable.
While we assume that it is applicable to Hedera, this might not be a justifiable assumption for other permissioned settings.
The applicability of this model to other permissioned systems should therefore be more formally analyzed.

\section{Conclusion}

The increasing popularity of \ac{DLT} systems since the invention of Bitcoin, and with it the energy-intensive \ac{PoW} consensus mechanism, has produced a variety of alternative mechanisms.
\ac{PoS} is a particularly popular alternative that is commonly assumed to be more energy efficient than \ac{PoW}.
In this paper, we tested this hypothesis using a mathematical consumption model that predicts expected energy consumption per transaction, as a function of network load.
Applying this model to six different \ac{PoS}-based \ac{DLT} systems supports the hypothesis and suggests that their energy consumption per transaction is indeed at least two to three orders of magnitude lower than that of Bitcoin.
Furthermore, we discover significant differences among the analyzed \ac{PoS}-based systems themselves.
Here, a permissioned system was found to consume significantly less energy per transaction than permissionless systems.
This difference could be attributed to gatekeeping capabilities offered by permissioned systems.

These results can be understood as an urgent call for the modernization of \ac{PoW} systems and a shift towards \ac{PoS}, as well as a recommendation to practitioners to consider appropriate, energy-saving hardware.
They are also intended to provide a basis for the future comparative study of the energy friendliness of \ac{PoS} systems and to facilitate the development of more rigorous consumption models.
Given the enormous challenges posed by climate change, avoiding unnecessary energy consumption needs to be a high priority.
Our work shows that \ac{PoS}-based systems can contribute to this and could even undercut the energy needs of traditional central payment systems, raising hopes that \ac{DLT} can contribute positively to combatting climate change.

Future research should further develop and confirm these initial findings by improving the sophistication of the model and considering factors beyond network throughput, that may influence validator count.
It should, furthermore, consider the network-wide energy consumption beyond validator nodes (i.e., by including all full nodes and auxiliary services) to arrive at a more holistic view of the overall energy consumption of \ac{DLT} systems.
Applying benchmarking frameworks~\cite{Sedlmeir2021} to measure the actual energy consumption might be particularly worthwhile in the context of permissioned systems that aim for high performance.
In addition, analyzing the actual hardware configurations, instead of relying on rough estimates, might prove a worthwhile extension.
Finally, future work should assess the effects of moving from a permissioned to a permissionless model.

\section*{Acknowledgements}

We thank Michel Zade for comments that greatly improved the manuscript.

M.P. was supported by the University College London Centre for Blockchain Technologies. M.P. was also supported by Google Cloud via the Google Cloud Research Grant program.
D.P. was supported by the Engineering and Physical Sciences Research Council [EP/L015234/1], the EPSRC Centre for Doctoral Training in Geometry and Number Theory (The London School of Geometry and Number Theory), University College London, and by Imperial College London.

\section*{Author Contributions}

Conceptualization: M.P., J.X., P.T., N.V. and J.I.I.; 
Data curation: M.P., J.S. and D.P.;
Formal analysis: D.P.;
Investigation: M.P., J.S., D.P., J.X and J.I.I.; 
Methodology: M.P., J.S. and D.P.; 
Visualization: M.P. and D.P.;
Writing -- original draft: M.P.; 
Writing -- review \& editing: M.P., J.S., D.P., U.G., J.X., P.T., N.V. and J.I.I..

\section*{Conflict of Interest}

M.P. declares that he is bound by a confidentiality agreement that prevents him from disclosing his competing interests in this work. 

\printacronyms

\printbibliography


\raggedbottom

\vspace{1em}

\appendix

\subsection{Validator Metrics}
\label{appdx:validators}

\begin{table}[H]
    \centering
    \begin{tabularx}{\columnwidth}{LLLrr}
        \toprule
        \textbf{Chain} & \textbf{Source} & \textbf{Metric} & \textbf{Obs. Period} & \textbf{Value} \\
        \midrule
        \textbf{Ethereum 2.0} & \url{https://beaconcha.in/charts} & Number of active validators & 5/7/2021 & \num[scientific-notation=false,round-mode=places]{\NumValEth} \\
        \textbf{Algorand} & \url{https://metrics.algorand.org/} & Number of nodes & 12/8/2021 & \num[scientific-notation=false]{\NumValAlgo} \\
        \textbf{Cardano} & \url{https://cardanoscan.io/} & Number of stake pools & 11/8/2021 & \num[scientific-notation=false]{\NumValCardano} \\
        \textbf{Polkadot} & \url{https://polkadot.subscan.io/validator} & Number of validators & 5/7/2021 & \num[scientific-notation=false]{\NumValPolka} \\
        \textbf{Tezos} & \url{https://tzstats.com/bakers} & Number of bakers & 12/8/2021 & \num[scientific-notation=false]{\NumValTezos} \\
        \textbf{Hedera} & \url{https://docs.hedera.com/guides/mainnet/mainnet-nodes} & Numbers of mainnet nodes & 13/8/2021 & \num[scientific-notation=false]{\NumValHed}s \\
        \bottomrule
    \end{tabularx}
    \vspace{0.1cm}
    \caption{Sources for data on contemporary validator machine count}
    \label{tab:current-validators}
\end{table}

\begin{table}[H]
    \centering
    \begin{tabularx}{\columnwidth}{LLLrr}
        \toprule
        \textbf{Chain} & \textbf{Source} & \textbf{Metric} & \textbf{Obs. Period} & \textbf{Value} \\
        \midrule
        \textbf{Polkadot} & \url{https://web.archive.org/web/*/https://stakers.info/} & Number of validators & 27/2/2021 & \num[scientific-notation=false]{\NumValPolkaHistoric}
        \\
        \textbf{Tezos} & \url{https://api.tzstats.com/explorer/cycle/324} & Number of bakers & 5/2/2021 & \num[scientific-notation=false]{\NumValTezosHistoric}
        \\
        \textbf{Algorand} & \url{https://metrics.algorand.org/} & Number of validators & 5/7/2021 & \num[scientific-notation=false]{\NumValAlgoHistoric} \\
        \textbf{Hedera} & \url{https://docs.hedera.com/guides/mainnet/mainnet-nodes} & Number of validators & 5/7/2021 & \num[scientific-notation=false]{\NumValHedHistoric} \\
        \textbf{Hedera} & \url{https://github.com/hashgraph/hedera-docs/commits/master/mainnet/mainnet-nodes/README.md} & Number of validators & 7/7/2020--26/8/2021 & \num[scientific-notation=false]{\NumValHedHistoric} \\
        \bottomrule
    \end{tabularx}
    \vspace{0.1cm}
    \caption{Sources for data on historic validator machine count}
    \label{tab:historic-validators}
\end{table}

\subsection{Throughput Metrics}
\label{appdx:throughput}

\begin{table}[H]
    \centering
    \begin{tabularx}{\columnwidth}{LLLRr}
        \toprule
        \textbf{Chain} & \textbf{Source} & \textbf{Metric} & \textbf{Obs. Period} & \textbf{Value} \\
        \midrule
        \textbf{Algorand} & \url{https://algoexplorer.io/} & Average transaction volume & 16/7/2021-12/8/2021 & \SI[scientific-notation=false]{\TpsAlgo}{\tx\per\second} \\
        \textbf{Cardano} & \url{https://explorer.cardano.org/en} & Number of transactions in epoch & Epoch 282 (3/8/2021-8/8/2021) & \SI[scientific-notation=false,round-mode=places]{\TxCardano}{\tx} \\
        \textbf{Polkadot} & \url{https://polkadot.subscan.io/extrinsic} & Mean of the lowest and the highest daily transaction volume & 5/6/2021-5/7/2021 & \SI[scientific-notation=false]{\TpsPolka}{\tx\per\second} \\
        \textbf{Tezos} & \url{https://tzstats.com/} & Average number of transactions per second & 13/7/2021-12/8/2021 & \SI[scientific-notation=false]{\TpsTezos}{\tx\per\second} \\
        \textbf{Hedera} & \url{https://hedera.com/dashboard} & Transaction volume by network service & 13/8/2021 & \SI[scientific-notation=false]{\TpsHed}{\tx\per\second} \\
        \bottomrule
    \end{tabularx}
    \vspace{0.1cm}
    \caption{Sources for data on contemporary throughput}
    \label{tab:current-throughput}
\end{table}

\begin{table}[H]
    \centering
    \begin{tabularx}{\columnwidth}{LLLRr}
        \toprule
        \textbf{Chain} & \textbf{Source} & \textbf{Metric} & \textbf{Obs. Period} & \textbf{Value} \\
        \midrule
        \textbf{Algorand} & \url{https://algoexplorer.io/} & Transactions per second & 2/6/2021-2/7/2021 & \SI[scientific-notation=false,round-mode=places, round-precision=1]{\TpsAlgoHistoric}{\tx\per\second} \\
        \textbf{Tezos} & \url{https://messari.io/asset/tezos} & Average number of transactions per second & 6/1/2021-5/2/2021 & \SI[scientific-notation=false,round-mode=places, round-precision=1]{\TpsTezosHistoric}{\tx\per\second} \\
        \textbf{Hedera} & \url{https://hedera.com/dashboard} & Transactions per second & 5/7/2021 & \SI[scientific-notation=false,round-mode=places, round-precision=1]{\TpsHedHistoric}{\tx\per\second} \\
        \textbf{Hedera} & \url{https://app.dragonglass.me/hedera/home} & Transactions per second & 8/2020--8/2021 & - \\
        \bottomrule
    \end{tabularx}
    \vspace{0.1cm}
    \caption{Sources for data on historic throughput}
    \label{tab:historic-throughput}
\end{table}

\begin{table}[H]
    \centering
    \begin{tabularx}{\columnwidth}{LLLr}
        \toprule
        \textbf{Chain} & \textbf{Source} & \textbf{Metric} & \textbf{Value} \\
        \midrule
        \textbf{Ethereum 2.0} & \url{https://twitter.com/VitalikButerin/status/1277961594958471168} & Transactions per second with Ethereum 1 as data layer & \SI[scientific-notation=false]{\TpsEthMax}{\tx\per\second} \\
        \textbf{Algorand} & \url{https://www.algorand.com/resources/blog/algorand-2021-performance} & Current maximum transactions per second & \SI[scientific-notation=false]{\TpsAlgoMax}{\tx\per\second} \\
        \textbf{Cardano} & \url{https://vacuumlabs.com/blog/lifevacuum/what-we-love-about-cardano-a-technical-analysis} & Maximum theoretical throughput & \SI[scientific-notation=false]{\TpsCardanoMax}{\tx\per\second} \\
        \textbf{Polkadot} & \url{https://twitter.com/gavofyork/status/1255859146127179782} & Sustained transactions per second & \SI[scientific-notation=false]{\TpsPolkaMax}{\tx\per\second} \\
        \textbf{Tezos} & \url{https://blockfyre.com/tezos-xtz/} & Transactions per second & \SI[scientific-notation=false]{\TpsTezosMax}{\tx\per\second} \\
        \textbf{Hedera} & \url{https://hedera.com/hbar} & Transactions per second & \SI[scientific-notation=false]{\TpsHedMax}{\tx\per\second} \\
        \bottomrule
    \end{tabularx}
    \vspace{0.1cm}
    \caption{Sources for data on maximum throughput}
    \label{tab:maximum-throughput}
\end{table}

\begin{table}[H]
    \centering
    \begin{tabularx}{\columnwidth}{LLLRr}
        \toprule
        \textbf{Bound} & \textbf{Source} & \textbf{Metric} & \textbf{Obs. Period} & \textbf{Value} \\
        \midrule
        \textbf{Lower} & \url{https://etherscan.io/} & Throughput of Ethereum~1 & 24/7/2021 & \SI[scientific-notation=false]{\TpsEthLow}{\tx\per\second} \\
        \textbf{Upper} & \url{https://twitter.com/VitalikButerin/status/1277961594958471168} & Postulated maximum transactions per second & - & \SI[scientific-notation=false]{\TpsEthHigh}{\tx\per\second} \\
        \bottomrule
    \end{tabularx}
    \vspace{0.1cm}
    \caption{Sources for throughput estimates for Ethereum 2.0}
    \label{tab:ethereum-estimates}
\end{table}

\subsection{Validator Energy Consumption}
\label{appdx:consumption}

\begin{table}[H]
    \centering
    \begin{tabularx}{\columnwidth}{LLLr}
        \toprule
        \textbf{Hardware} & \textbf{Source} & \textbf{Metric} & \textbf{Value} \\
        \midrule
        \textbf{Raspberry Pi 4} & \url{https://www.tomshardware.com/uk/reviews/raspberry-pi-4} & Power consumption when idle & \SI[scientific-notation=false,round-mode=places, round-precision=1]{\PRasPiLow}{\watt} \\
        \textbf{Raspberry Pi 4} & \url{https://www.tomshardware.com/uk/reviews/raspberry-pi-4} & Power consumption under load & \SI[scientific-notation=false,round-mode=places, round-precision=1]{\PRasPiHigh}{\watt} \\
        \textbf{Dell PowerEdge R730} & \url{https://i.dell.com/sites/csdocuments/CorpComm_Docs/en/carbon-footprint-poweredge-r730.pdf} & Typical yearly energy consumption & \SI[scientific-notation=false,round-mode=places, round-precision=1]{\TECServer}{\kilo\watt\hour} \\
        \textbf{Hewlett Packard Enterprise ProLiant ML350 Gen10} & \url{https://www.spec.org/power_ssj2008/results/res2019q2/power_ssj2008-20190312-00899.html} & Power consumption under 80\% load & \SI[scientific-notation=false,round-mode=places, round-precision=1]{\PHighEndServer}{\watt} \\
        \bottomrule
    \end{tabularx}
    \vspace{0.1cm}
    \caption{Sources for data on hardware energy consumption}
    \label{tab:hardware-consumption}
\end{table}

\subsection{Comparison Values}
\label{appdx:benchmark}

\begin{table}[H]
    \centering
    \begin{tabularx}{\columnwidth}{LLLRr}
        \toprule
        \textbf{System} & \textbf{Source} & \textbf{Metric} & \textbf{Obs. Period} & \textbf{Value} \\
        \midrule
        \textbf{Bitcoin} & \url{https://cbeci.org/} & Theoretical lower bound of annualized power consumption & 11/8/2021 & \SI{\AnnualisedConsumptionBitcoinLow}{\kilo\watt\hour}  \\
        \textbf{Bitcoin} & \url{https://cbeci.org/} & Theoretical upper bound of annualized power consumption & 11/8/2021 & \SI{\AnnualisedConsumptionBitcoinHigh}{\kilo\watt\hour} \\
        \textbf{Bitcoin} & \url{https://www.blockchain.com/charts/transactions-per-second} & Transactions per second & 30 day average on 11/8/2021 & \SI[scientific-notation=false]{\TpsBitcoin}{\tx\per\second} \\
        \textbf{VisaNet} & \url{https://usa.visa.com/content/dam/VCOM/global/about-visa/documents/visa-2020-esg-report.pdf} & Approximate total energy consumption of the Visa corporation & 2020 & \SI[scientific-notation=false]{\AnnualisedConsumptionVisa}{\giga\joule} \\
        \textbf{VisaNet} & \url{https://usa.visa.com/run-your-business/small-business-tools/retail.html} & Transactions per day & 8/2010 & \SI{\TpdVisa}{\tx\per\day}  \\
        \bottomrule
    \end{tabularx}
    \vspace{0.1cm}
    \caption{Sources for data on reference systems}
    \label{tab:reference-systems}
\end{table}

\end{document}